# Tunable, high pulse energy and narrow linewidth gas-filled fiber laser across near- and mid-infrared

Yazhou Wang, Marcello Meneghetti, Manoj K. Dasa, Timothy Bate, J. E. Antonio-Lopez, Ian A. Davidson, Qiang Fu, Jaroslaw Rzegocki, Gregory T. Jasion, Francesco Poletti, Rodrigo Amezcua-Correa, and Christos Markos

## Affiliations

**DTU Electro, Technical University of Denmark, Kgs. Lyngby, Denmark**

Yazhou Wang, Marcello Meneghetti, and Christos Markos

**NORBLIS ApS, Virum, Denmark**

Christos Markos

**NKT Photonics A/S, Birkerød, Denmark**

Manoj K. Dasa

**CREOL, The College of Optics and Photonics, University of Central Florida, Orlando, USA**

Timothy Bate, J. E. Antonio-Lopez, and Rodrigo Amezcua-Correa

**Optoelectronics Research Centre, University of Southampton, Southampton, UK**

Ian Davidson, Qiang Fu, Jaroslaw Rzegocki, Gregory T. Jasion, and Francesco Poletti

## Contributions

YW, MM and CM conceived the idea. YW designed the experiment and characterized the laser, supported by MM and MKD. TB, JEA-L, and RAC designed, fabricated, and characterized ARHCF-1, -2, and -4. ID, QF, JR, GTJ, and FP designed, fabricated, and characterized ARHCF-3 and -5. MM and YW analysed, plotted, and prepared the manuscript figures. The first version of the manuscript was prepared by YW, MM, and CM, while all the authors contributed to the final version.

## Corresponding author

Correspondence to: Yazhou Wang & Christos Markos

## Abstract

Wavelength widely tunable infrared fiber lasers that simultaneously deliver high pulse energies with narrow linewidths are critical for applications ranging from spectroscopy to nonlinear optics, yet achieving this combination has remained a long-standing challenge. Here, we demonstrate that gas-filled anti-resonant hollow-core fiber Raman laser offers tunability across a broad spectral range from the near-infrared (~1.4 µm) to the mid-infrared (~4.6 µm), with near microjoule level high pulse energy and a narrow linewidth of few gigahertz or less. This performance arises from a unique pump laser design together with an optimized selection of gas-filled anti-resonant hollow-core fibers, opening a promising pathway toward compact, high-performance tunable infrared fiber-based sources.

## Introduction

Infrared lasers that simultaneously provide narrow linewidth, high pulse energy, and broad wavelength tunability across the near- and mid-infrared regions play an essential role for trace-level multi-gas sensing [1–3], multi-spectral photoacoustic microscopy [4,5], and nonlinear frequency conversion [6,7]. The dominant approaches for generating lasers with such characteristics are based on optical parametric frequency conversion, including optical parametric oscillators and amplifiers [8–12]. However, these laser technologies require free-space architectures incorporating nonlinear crystals and precision alignment, which compromises their mechanical stability and footprint. External-cavity semiconductor lasers, including external-cavity quantum cascade lasers [13–15], provide an alternative route. Yet, in addition to requiring free-space elements for wavelength tuning, their pulse energy in pulsed operation remains limited—typically to the sub-nanojoule level for tens of nanosecond pulse duration [16,17].

Monolithic all-fiber laser architectures constitute a promising way to overcome these challenges. Near-infrared all-fiber lasers with narrow linewidth and tunable wavelength have been extensively developed in the past decades [18–22]. This momentum has recently extended into the mid-infrared region using soft-glass fibers [23,24]. However, their wavelength tunability is fundamentally constrained to several tens of nanometres by the narrow gain bandwidth of active lasing media. Gas-filled silica anti-resonant hollow-core fibers (ARHCFs) have recently emerged as a promising alternative, offering broadband transmission from the ultraviolet to the mid-infrared [25–28], together with exceptionally high optical damage thresholds [29–31]. With these properties, gas-filled ARHCF lasers with high-energy (power) and narrow-linewidth have been intensively reported [32–39], and the recent advances in low-loss fusion splicing [31,40–44] suggest a future path toward all-fiber light sources. Nevertheless, wavelength tunability of gas-filled ARHCF lasers yet remains rarely explored. Two mechanisms can in principle enable the wavelength tunability. The first relies on the population inversion of active gas which can provide tuning over hundreds of nanometers, but only in discrete steps set by molecular absorption lines [44–48]. The second is stimulated Raman scattering (SRS) in gas, wherein the Raman-shifted wavelength can be controlled via the pump wavelength. However, all reported SRS-based ARHCF lasers have achieved tuning ranges of only a few tens of nanometres, far below what is required for broadband multispectral applications [32,49,50].

In this work, we demonstrate broad Raman wavelength tunability in gas-filled ARHCF lasers, spanning from the near- to mid-infrared spectral region. This wide wavelength range stems from a narrow tuning range of the pump wavelength, from 1034 nm to 1085 nm. The resulting Raman lasers maintain (sub-)GHz linewidths and near microjoule-level pulse energies across the entire spectral range. This performance is enabled by the combination of an optimized pump design and appropriate selection of gas-filled ARHCFs. Although our experimental implementation is partially based on free-space coupling, the underlying approach is fully compatible with all-fiber integration, supported by recent progress in single-frequency, wavelength-tunable near-infrared all-fiber pumps[18–22], and low-loss splicing techniques for ARHCFs [31,40–44].

## Concept & major results

Figure 1a illustrates the overall concept. A wavelength-tunable near-infrared pump laser (1034–1085 nm) is coupled into a two-stage cascade of gas-filled ARHCFs. In the 1st stage ARHCF, the pump is converted into a wavelength-tunable near-infrared Raman Stokes signal, which then acts as the pump for the 2nd stage ARHCF to generate a Raman output spanning a substantially extended infrared range.

**Fig. 1: Concept illustration & main results.**

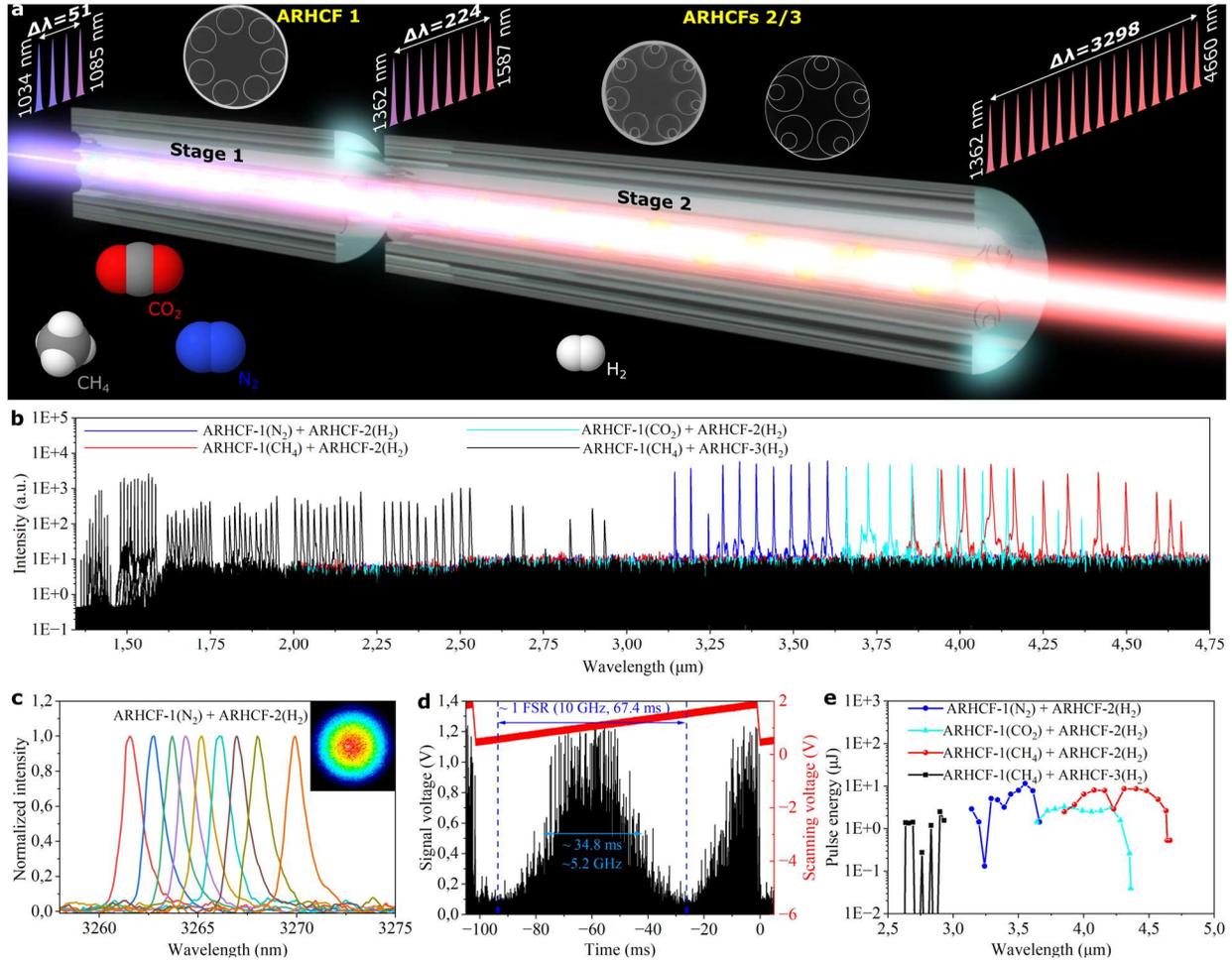

**a** Concept of the proposed tunable gas-filled Raman laser. A pump laser tunable from 1034 to 1085 nm drives the cascaded two-stage gas-filled ARHCFs, enabling tunable Raman lines spanning ~1.4–4.6 μm. **b** Measured Raman spectra from two parallel 2$^{nd}$ stage ARHCFs (ARHCF-2 and ARHCF-3) with pump wavelengths at 1034, 1040, 1045, … 1085 nm; the 1$^{st}$ stage ARHCF is denoted as ARHCF-1. The two longest-wavelength lines (in red-color) correspond to pump wavelengths of 1083 and 1084 nm. The complete spectra are provided in Fig. S1. **c** An example of continuous wavelength tuning near 3.2 μm; inset image shows the beam profile recorded with a mid-infrared beam profiler (S-WinCamD-IR-BB-7.5, DataRay). **d** Linewidth of a 4.01 μm Raman line measured using a scanning Fabry–Perot interferometer (SA210-30C-SP, Thorlabs). The black curve shows the interferometric envelope of the Raman line; the red trace indicates the applied scanning voltage. This 4.01 μm Raman line was generated through cascading $CH_4$-filled ARHCF-1 at 2 bar and $H_2$-filled ARHCF-2 at 30 bar. **e** Pulse energies of tunable Raman lines in the mid-infrared region. The symbol '+' in the legends of **b** and **e** denotes cascading two ARHCFs, with gas species indicated in brackets. ARHCF: anti-resonant hollow-core fiber. FSR: free spectral range.

The underlying motivation of this configuration is to exploit the dependence of gas-based Raman wavelength tunability on that of the Yb-doped pump laser. The Raman Stokes wavelength directly follows the pump wavelength tuning, and its tuning range is inherently broader than that of the pump [32]. Moreover, the Raman wavelength tuning range increases with the Raman frequency-shift coefficient and scales inversely with the pump wavelength [32]. These considerations guided two key design choices: (i) A Yb-doped fiber laser was chosen as the pump source, as it operates at the shortest achievable wavelength (~1 µm) among commonly used rare-earth–doped fiber lasers, and (ii) two gas-filled ARHCFs are cascaded to effectively increase the overall Raman frequency shift coefficient.

The pump laser is designed to emit a narrow spectral line tunable from 1034 to 1085 nm, which is well within the gain range of Yb-doped fibers. Its linewidth is ~5 GHz, and it delivers ~3.4 ns pulses at a 1 kHz repetition rate with pulse energies of 70–100 µJ across the tuning range. To clarify the achievable Raman performance, we set aside the detailed pump design and first focus on the wavelength ranges obtained from the 2$^{nd}$ stage ARHCFs. Three ARHCFs are used in this part — denoted ARHCF-1, ARHCF-2, and ARHCF-3 (see Supplementary Note 1). ARHCF-1 serves as the 1$^{st}$ stage Raman converter, while ARHCF-2 and ARHCF-3 function as parallel 2$^{nd}$ stage fibers. The Raman output of ARHCF-1 can be coupled into either ARHCF-2 or ARHCF-3 through a flip mirror. The details of the cascaded ARHCF configuration are provided in Method 1.

In the configuration of cascading ARHCF-1 and ARHCF-2, ARHCF-1 is selectively filled with one of three gases including $N_2$, and $CO_2$, and $CH_4$, to enable their 1$^{st}$/2$^{nd}$ order vibrational Stokes generation with Raman frequency shift coefficients of 2331 cm$^{-1}$ ($v_1$ mode of $N_2$) [51], 1385 cm$^{-1}$ ($v_1$/2$v_2$ mode of $CO_2$) [52], and 2917 cm$^{-1}$ ($v_1$ mode of $CH_4$) [38], respectively. Gas exchange is achieved by purging the existing gas with the selected gas. The pressure is set to 15 bar for $N_2$, 2 bar for $CO_2$, and 2 bar for $CH_4$, to optimize their Raman conversion efficiencies. With these three gases, the wavelength tunable pump can be converted to Raman in the near-infrared region of 1362-1587 nm, where each gas oversees a portion of this wavelength range (see Fig. S2a- S2c). The linewidth of the Raman is measured to be less than ≲ 7 GHz (see Fig. S2d-S2e). The Raman wavelength is continuously tunable in a wavelength range that is misaligned with any strong absorption bands of the selected gas. High-order Raman Stokes with wavelength > 1.6 µm is well suppressed, due to the tailored loss profile of ARHCF-1 (see Supplementary Note 1). As a result, high quantum efficiency in the range of 30%-86% and thus high pulse energy of 18-48 µJ are achieved (see Fig. S2g-S2h), which is sufficient to enable the SRS process of the 2nd stage $H_2$-filled ARHCF-2.

The 1st stage ARHCF-1 Raman laser is coupled into the 2nd stage ARHCF-2. To enable a broad wavelength tunability, the selected ARHCF-2 has two transmission windows, respectively at the pump wavelength region (1362 nm-1587 nm), and the mid-infrared region (~3 - 4.6 µm). $H_2$ is selected as the Raman active gas, to take its unique advantages of the large Q(1) vibrational Raman frequency shift coefficient of 4155 cm$^{-1}$, high Raman gain coefficient [53], and negligible absorption effect across the near- and mid-infrared regions [53]. At 30 bar $H_2$ pressure, ARHCF-2 Raman lasers with a broad wavelength tunable range from 3140 nm to 4660 nm are converted from the 1362 nm-1587 nm Raman wavelength range of the ARHCF-1. Figure 1b shows the tunable Raman spectral lines measured with a spectrum meter (Spectro 320, Instrument Systems), where three groups of Raman lines based on the three gases ($N_2$, $CO_2$, and $CH_4$) filled in the 1st stage ARHCF-1 are denoted in different colors, to distinguish each's wavelength range. Specifically, their wavelength ranges are 3.14-3.66 µm, 3.64-4.36 µm, and 3.85-4.66 µm, respectively, when pumped by the 1st order Raman Stokes of $N_2$, 2nd order Stokes of $CO_2$, and 1st order Stokes of $CH_4$ filled in ARHCF-1. Note that the 1st order $CO_2$ vibrational Raman Stokes lines (1207-1277 nm) are blocked by a 1.3 µm long pass filter from being coupled into ARHCF-2, since ARHCF-2 has a relatively high loss at this wavelength range and its corresponding vibrational Raman Stokes of $H_2$ (2.42-2.72 µm).

The second parallel configuration employs $H_2$-filled ARHCF-3 as the 2nd stage ARHCF, to bridge the wavelength gap from ~1.4-3 µm region. In this configuration, ARHCF-1 and ARHCF-3 are filled with $CH_4$ at 2 bar and $H_2$ at 30 bar, respectively. With the properly tailored loss profiles of ARHCF-1 and ARHCF-3, this configuration combines the 1st order $v_1$ mode vibrational Stokes generation of $CH_4$, and S(1) mode high order rotational Stokes generation of $H_2$. Their Raman frequency shift coefficients are 2917 cm$^{-1}$ and 587 cm$^{-1}$, respectively. The black-color spectra in Fig. 1b show the measured results, where up to 5th order cascaded rotational Raman Stokes lines and 1st order rotational Raman anti-Stokes are generated from the $H_2$-filled ARHCF-3, covering most spectral regions from ~1.4 to 3 μm. This broad Raman wavelength range also benefits from the suppression of 1$^{st}$ order Q(1) vibrational Raman Stokes generation of $H_2$, which has its wavelength at the 4 μm high loss region of the ARHCF-3. Due to the limited pump wavelength tuning range, narrow wavelength gaps are formed between the adjacent wavelength tuning ranges of each order of S(1) rotational Raman (anti-)Stokes (pump can be denoted as 0 order Stokes) of $H_2$. At the wavelength range of the 5$^{th}$ order Stokes, the Raman line tunability becomes discontinuous, because of the waveguide loss introduced by the non-uniform capillary thickness of ARHCF-3.

In addition to its broad wavelength tuning range, the developed Raman laser exhibits continuous wavelength tunability with a narrow linewidth of a few GHz. Figure 1c illustrates an example of continuous tuning over ~10 nm around the 3.2 μm region, based on cascaded $N_2$-filled ARHCF-1 and $H_2$-filled ARHCF-2. The tuning step is approximately 1 nm, although significantly finer precision is allowed (see 'Discussion and outlook'). The measured linewidth of ~1 nm shown in Fig. 1c is limited by the spectrometer resolution (0.4 nm; Spectro 320, Instrument Systems) and the use of a multimode fiber patch cable (MF11L1AR1, Thorlabs) for coupling. Accurate linewidth measurements were performed using Fabry–Perot interferometers (see Method 2). Figure 1d shows a measured linewidth of ~5 GHz (~0.27 nm) for the Raman laser at 4.01 μm based on cascading $CH_4$-filled ARHCF-1 and $H_2$-filled ARHCF-2. The selection of 4.01 μm takes advantage of the atmospheric transmission window, to avoid the absorption impact of ambient air on the linewidth measurement. Linewidths of other Raman lasers based on different configurations of cascaded ARHCFs are provided in Fig. S3, where they are in the range of 4-8 GHz. These narrow linewidth values are mainly attributed to three factors: the narrow linewidth of the pump, the narrow gain bandwidth of gas in few GHz or less, and the spectral gain narrowing effect of SRS. Furthermore, all Raman lasers in this study exhibit a Gaussian-like beam profile, exemplified by the 3.2 μm Raman beam in the inset of Fig. 1c, owing to the lower loss of the fundamental mode compared to higher-order transverse modes in the employed ARHCFs.

Across the above Raman wavelength tuning ranges, pulse energies can be basically maintained at the microjoule level, with few nanoseconds pulse width. Figure 1e shows the measured pulse energy in mid-infrared region. When the wavelength is ≳ 4.6 µm, the Raman energy is limited by the high silica absorption loss of ARHCF-2. Note that the Raman pulse energy can experience a significant propagation attenuation in the ambient air, when its spectral line coincides with one of the strong absorption lines of atmospheric gases. For example, in the Supplementary Note 2, our measurements show that by precisely tuning the Raman line at around 4.23 µm, a high attenuation coefficient of 8.1 m$^{-1}$ was observed due to $CO_2$ absorption, which approaches the calculated peak $CO_2$ absorption coefficient of 12 m$^{-1}$ at the ambient concentration of ~400 ppm. Due to the narrow linewidth, this attenuation coefficient is also ~6 times higher than that in our previous work [32], indicating the potential of the proposed Raman laser on trace level gas spectroscopy.

## Design of pump laser

The broad wavelength tunability, narrow linewidth, and high pulse energy of the proposed Raman laser critically depend on those of the pump laser. Figure 2a shows the configuration of the pump design, mainly consisting of a linearly polarized continuous wave (CW) laser seed (TEC-500-1060-030, Sacher Lasertechnik), followed by an $LiNbO_3$ phase modulator (NIR-MPX-LN-02, Exail Technologies), an semiconductor optical amplifier (SOA) which serves as an intensity modulator (1020SOA-1-5-1, Aerodiode), and a Yb-doped fiber amplification module. Yb-doped fiber lasers have been extensively developed in terms of wide wavelength tunability, narrow linewidths [54,55], high pulse energy [56–58]. The objective of the pump design is to simultaneously combine these three metrics.

The wavelength tunability and narrow linewidth properties of the pump laser stem from the CW laser seed, which is an external cavity diode laser, emitting a linearly polarized spectral line with 10 kHz narrow linewidth, tunable in the custom range of 1000 nm to 1085 nm. Through the SOA intensity modulator, the CW seed is modulated into a nanosecond pulse train operating at 1 kHz repetition rate, where the pulse width and pulse shape can be controlled with the SOA intensity modulator. The temporal profile of the pulse is tailored to an exponential profile, to pre-compensate gain distortion introduced during the optical amplification. The fiber amplification module has a polarization-maintained fiber structure, consisting of

**Fig. 2: Design and characterization of the pump laser.**

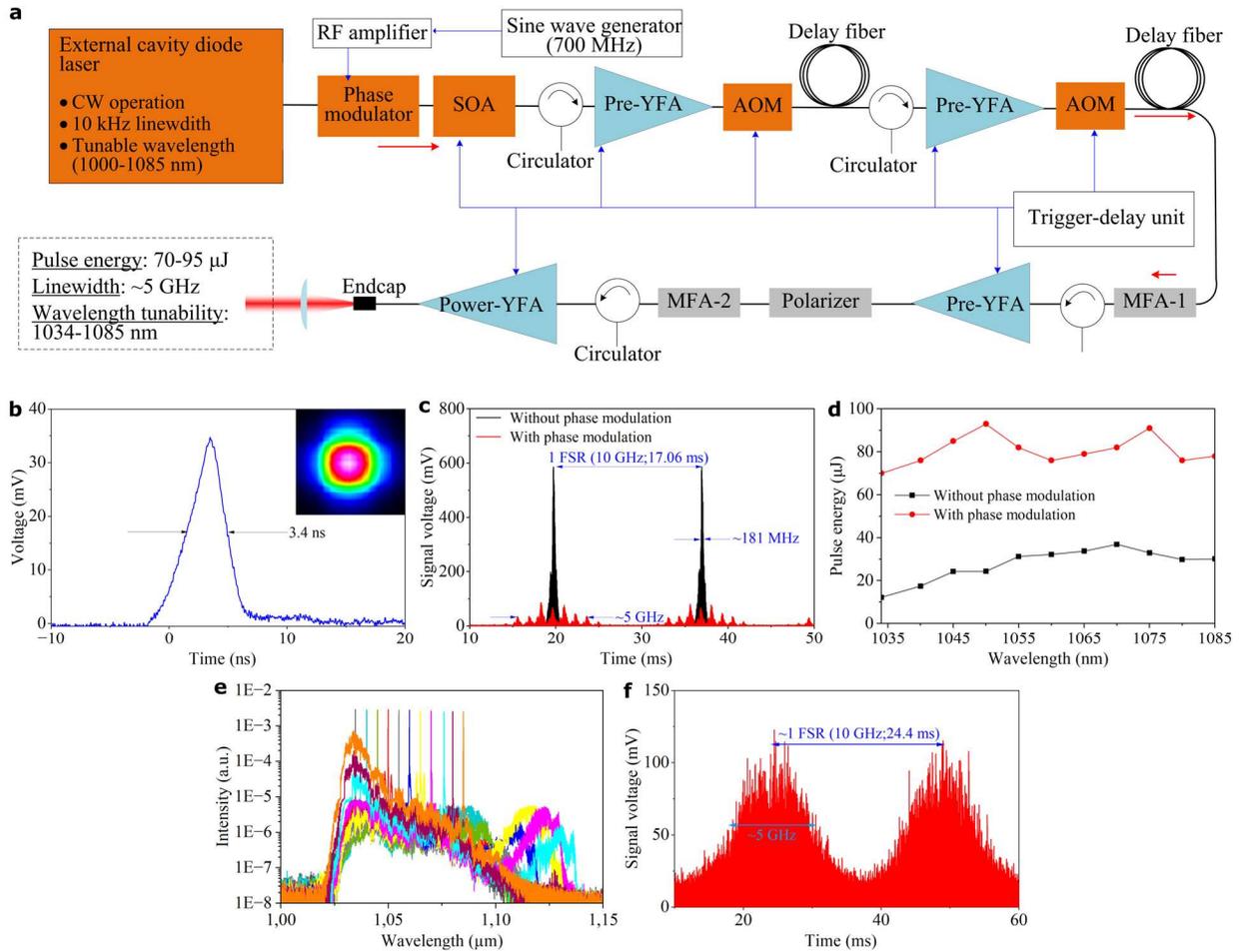

**a** Experimental configuration of the pump laser. **b** Measured pulse profile at 1060 nm with 75 μJ pulse energy; inset shows the corresponding beam profile. **c** Linewidth measurement of the 3.4 ns seed laser using a scanning Fabry–Perot interferometer (SA210-8B, Thorlabs) with a free spectral range (FSR) of 10 GHz and resolution of 67 MHz. Red and black traces correspond to measurements with and without phase modulation, respectively. **d** Pulse energies of the wavelength-tunable pump laser with and without phase modulation, measured using an energy meter (PE9-ES-C, Ophir Optronics). **e** Spectra of the wavelength tunable and high energy pump laser (with phase modulation), measured with a spectrum analyzer (ANDO AQ6317B, AssetRelay, resolution: 10 pm). **f** Linewidth measurement of pump pulse at 1060 nm, with a pulse energy of 75 μJ. The trace in **f** is recorded with a scanning Fabry-Perot interferometer (SA210-8B; Thorlabs). SOA: semiconductor optical amplifier. AOM: acoustic optical modulator. Both SOA and AOMs play the role of intensity modulation. YFA: Yb-doped fiber amplifier. RF amplifier: radio frequency amplifier. MFA-1: mode-field adaptor from PM980 (Coherent) to PLMA-10/125-M (Coherent) fiber. MFA-2: mode-field adaptor from PLMA-10/125-M to PLMA-25/250-M (Coherent) fiber.

three stages Yb-doped pre-amplifiers and a Yb-doped power amplifier. While the details of the pump laser are provided in Method 3, here we highlight the key aspects of the design. i) The pulse width is optimized to ~3 ns (see Fig. 2b). This pulse width is much shorter than the dephasing time of ~10 ns of silica-based stimulated Brillouin scattering (SBS), while far longer than the dephasing time of ≲1 ns of gas-based SRS involved in this experiment. Therefore, this selected pulse width can suppress detrimental Brillouin lasers through introducing transient SBS, while the gas-based SRS effect operates at a quasi-steady-state with a high Raman gain. ii) We further suppress the SBS effect through externally applying a sinusoidal optical phase modulation on the seed laser. With this measure, the linewidth of the seed pulse train is properly broadened from ~180 MHz to ~5 GHz through the phase modulator (see Fig. 2c). The frequency of the sinusoidal phase modulation is optimized to 700 MHz, as additional enhancement in SBS suppression is minimal by further increasing the modulation frequency [59]. iii) Tuning the pump wavelength across 1034 to 1085 nm is associated with a significant amount of amplified spontaneous emission (ASE), particularly when the wavelength is tuned away from the gain center region of the amplification module. To temporally suppress the ASE, two fiber-coupled acousto-optic intensity modulators (AOMs) are placed after the 1st and 2nd stage Yb-doped fiber amplifiers. In addition, these two AOMs also play the role of a broadband isolator, to effectively block the detrimental back-propagation laser pulses, particularly SBS lasers. This is achieved by adding a few meters long 'delay fiber' to the output end of the AOM, so that the AOM switches to the 'off' state when the backward propagating pulses arrive. iv) Given the fact that the gain range shifts towards long-wavelength by increasing the length of Yb-doped fiber, the Yb-doped fibers used in the 1st and 2nd stage pre-amplifier are tailored to 6 m and 3 m, respectively, to improve the bandwidth and flatness of the amplification gain profile [60].

With the design above, the pump energy in the range of 70 to 95 µJ is achieved in the wavelength range of 1034-1085 nm (see Fig. 2d and 2e). The pulse energy is measured with an energy meter (PE9-ES-C, Ophir Optronics), excluding the influence of ASE (see Fig. S4). When the phase modulation is off, the pulse energy is lowered to the range of 15-40 µJ (see Fig. 2d). The maximum pulse energy is limited by the backward propagation SBS lasers, which is monitored through optical circulators placed before each stage Yb-doped amplifier (see Fig. 2a). Further amplifying the pulse energy induces excessive SBS light, which might beyond the damage threshold of the amplification module. The linewidth of the pump laser is measured to be ~ 5 GHz (see Fig. 2f), which is approximately consistent with the linewidth of phase modulated seed laser at Fig. 2c. Compared with Fig. 2c, the absence of phase modulation structure in Fig. 2f is attributed to spectral broadening induced by self-phase-modulation effect in the Yb-doped fiber amplification module.

## Discussion and outlook

A natural question arising from this work is whether the Raman linewidth can be further reduced to the MHz level. We investigated this possibility through further narrowing the pump linewidth, which is achieved by turning off the phase modulation of the seed laser of the pump (see Fig. S5a – S5c), and by appropriately prolonging the pump pulse width through setting the SOA modulator. These two measures are, however, at the expense of lowering the pulse energy (see Fig. 2d), due to the decreased threshold of SBS. For example, at ~5 ns pulse width and without the phase modulation, the pump linewidth is reduced to ~180 MHz, but associated with a lowered pulse energy of 6-25 μJ in the wavelength tuning range of 1034-1085 nm (see Fig. S5d). Given the lowered pulse energy, cascaded $CO_2$-filled ARHCF-4 and $H_2$-filled ARHCF-5 (see Supplementary Note 1) are selected to implement the mid-infrared Raman frequency conversion at 2.5 μm region. The tailored low-loss transmission window of ARHCF-4 allows the Raman frequency conversion of the pump wavelength range to 1207-1277 nm through the $1^{st}$ order vibrational $CO_2$ Raman Stokes generation, while blocks the high order vibrational Raman Stokes through the high fiber loss above 1.3 μm wavelength (see Fig. 3a). The Raman pulse energy is in the range of 1.3 to 11.7 μJ (see Fig. S5e). Figure 3b shows the measured linewidth of the 1228 nm Raman laser at different pump pulse widths, where the Raman linewidth is reduced from 657 MHz at 3.4 ns to 384 MHz at 5.7 ns and to 197 MHz at 20.2 ns. Note that the time coordinates of the signal peaks in Fig. 3b show a slight drift at different measurements, which is caused by the frequency drift of the scanning Fabry-Perot interferometer due to the variation of ambient temperature.

At the pump pulse duration of ~5.7 ns, mid-infrared wavelength tunable Raman laser with ~700 MHz narrow linewidth was observed by coupling the $CO_2$-filled ARHCF-4 Raman laser into the $2^{nd}$ stage $H_2$-filled ARHCF-5 through the Q(1) vibrational SRS of $H_2$, as show in Fig. 3a. The longest measured Raman

**Fig. 3: Wavelength-tunable gas-filled fiber Raman laser with MHz-level linewidth.**

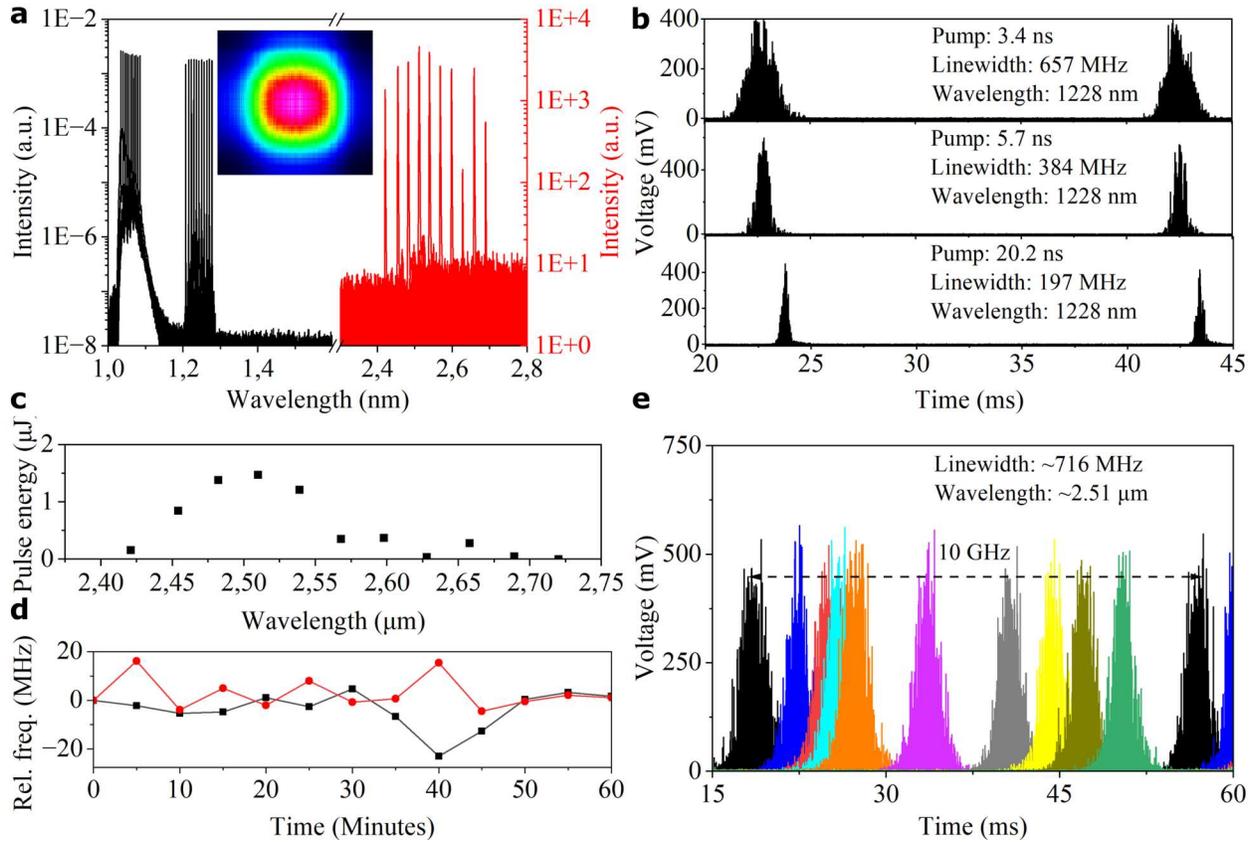

**a** Output spectra from the 1st stage $CO_2$-filled ARHCF-4 (black color) and the 2nd stage $H_2$-filled ARHCF-5 (red color) through tuning the pump wavelength to 1034, 1040, 1045, ... 1085 nm. The complete spectra in black-color are provided in Fig. S5f. **b** Linewidth measurement of the 1228 nm Raman laser generated in $CO_2$-filled ARHCF-4 when pumped at 1050 nm with different pulse durations. **c** Mid-infrared Raman pulse energy measured at the output of the 2nd stage $H_2$-filled ARHCF-5. **d** Frequency stability of the 1228 nm near-infrared Raman line. The red and black data are extracted from different peaks of the scanned Fabry–Perot interferometric trace (see Method 4). **e** Fine wavelength tuning of the mid-infrared Raman output from the 2nd stage $H_2$-filled ARHCF, at around 2.5 μm wavelength.

wavelength is 2.69 μm, converted from the 1080 nm pump line. The complete spectra of Fig. 3a are provided in Fig. S5f as a reference. Figure 3c shows the measured pulse energy of the mid-infrared Raman laser, where microjoule pulse energy is achieved around 2.5 μm wavelength. < 1 μJ pulse energy at other wavelengths is caused by the insufficient pump pulse energy and/or the increased loss of ARHCF-5 toward the mid-infrared region. Possible solutions to improve the pump/Raman pulse energy include optimizing the suppression of SBS effect in the optical amplification stage of the pump, and lowering the loss of ARHCFs. Further suppression of SBS can be achieved by increasing the mode-field area of Yb-doped fiber by introducing the photonic crystal structure [61], and by optimizing the waveform of the radio frequency signal applied to the optical phase modulation [62,63].

With the MHz narrow linewidth, another practical concern is the wavelength stability of the gas-filled Raman laser. Here we measured the wavelength stability of the 1228 nm Raman with ~380 MHz linewidth (see Fig. 3b). This measurement is implemented by measuring its frequency difference to the pump via scanning Fabry-Perot interferometric trace, which includes the signals of both the pump and Raman (see Method 4). The wavelength stability is measured at 1228 nm instead of the mid-infrared region, because no commercial Fabry-Perot interferometer can maintain a high resolution (< 100 MHz) across near- and mid-infrared regions, to our best knowledge. Figure 3e shows the measured results over one hour, where the fluctuation of the measured Raman wavelength is less than ± 20 MHz. This excellent wavelength stability is attributed to two factors. i) The wavelength of the pump laser is well stabilized due to precise temperature control of the commercial seed laser. ii) The gas-filled ARHCFs remained at room temperature and constant pressure due to the small amount of heat generation, therefore avoiding the potential drift of Raman frequency shift coefficient over time [64]. The heat generation mainly originates from the fiber loss and the Raman quantum defect. With $CO_2$-filled ARHCF-4, the fiber losses at both pump and Raman wavelengths are less than 1 dB/km, and the heat generation from the Raman quantum defect is as low as 2 mW. In the case of a high amount of heat release, the temperature of gas-filled ARHCFs can be controlled with cooling measures to stabilize the Raman wavelength.

With the ~700 MHz narrow Raman linewidth and the stable Raman wavelength, we observed the wavelength tunability of the Raman laser with a precision higher than that in Fig. 1d. For example, Fig. 3e shows the precise tuning of the 2.5 μm Raman over 10 wavelengths across the 10 GHz free spectral range of a Fabry-Perot interferometer. For the adjacent two Raman lines in Fig. 1d, the wavelength of pump/seed laser is tuned through the software-based automation control. Note that the Raman wavelength tuning step is not uniform in Fig. 3e, which is anticipated to be caused by the mode-hopping of the diode laser seed of the pump.

Another known but as-yet unrealized advantage of the proposed method is that the Raman Stokes wavelength can be tuned substantially faster than the pump. Achieving a fast wavelength tuning ability is critical for rapid, multi-spectral imaging and spectroscopy [3,65]. With the standard seed tuning speed of 200 nm/s in the 1 μm region, the corresponding Raman wavelength in the 4 μm region could, in principle, be tuned at an accelerated rate of ~3000 nm/s (see Supplementary Note 3). However, this capability was not demonstrated in this work, as the amplification parameters of the pump currently require manual adjustment for different wavelengths (Method 3). We anticipate that automated tuning of the Raman laser can be realized by synchronizing the amplification module with the pump wavelength sweep.

## Methods

### Method 1: Configuration of cascaded gas-filled ARHCFs

Each fiber in the cascaded ARHCF configuration is housed in a gas hose coiled to a diameter of ~30 cm, with both ends sealed by house-made gas cells. Gas is filled into and released from the ARHCF through the gas ports of gas cells. The design of the gas cell is provided in Fig. S6. The gas system can sustain a long-term stable pressure in the range of 1-30 bar. Laser coupling into and out of the ARHCF is achieved via optical windows mounted on each gas cell. To suppress detrimental back-reflection of 1 μm light into the pump laser, the two windows equipped on the gas cells of the 1st stage ARHCF employed nanostructured antireflective surfaces (W4105FT1, Thorlabs; <0.25% reflection @ 1 μm wavelength region). At the output side of the 1st stage ARHCF, a long pass filter is used to block the residual pump.

Given the output Raman Stokes laser from the 1st stage ARHCF always lies within 1.2-1.6 μm region, the input window of the 2nd stage AHRCF employed a C-coating (WG41050-C, Thorlabs), to minimize the transmission loss. The output window of the 2nd stage AHRCF is an uncoated $CaF_2$ window. Laser coupling was performed using plano-convex lenses, achieving ~80% coupling efficiency.

This method applies to all gas-filled cascaded ARHCFs used in this work.

### Method 2: Laser linewidth measurements

Three scanning Fabry-Perot interferometers (SA210-8B; SA210-18C; SA210-30C-SP; Thorlabs) are used to measure the laser linewidths across different wavelength ranges. Each interferometer provides a free spectral range (FSR) of 10 GHz and a resolution of 67 MHz at the specified wavelength range. A plano-convex uncoated $CaF_2$ lens with 10 cm focus length is used to couple the collimated laser beam into the selected Fabry-Perot interferometer. SA210-8B and SA210-18C are equipped with integrated photodetectors, which transmit data to an oscilloscope (MSO64B, Tektronix) operating in the envelope mode to obtain clear scanning traces at 1 kHz laser repetition rate. For the SA210-30C-SP Fabry-Perot interferometer, a HgCdTe amplified photodetector (PDAVJ8, Thorlabs) is attached to its output end. Neutral density filters are placed in front of Fabry-Perot interferometers, to appropriately attenuate the laser.

The scanned interferometric peaks show drift over time due to ambient temperature variations, complicating precise measurements of linewidth, wavelength tunability, and stability. To mitigate this effect, the interferometers were properly isolated from external heat sources using aluminium screens.

**Method 3: Design of pump laser**

The basic pump configuration is shown in Fig. 2a. The seed source is a commercial external-cavity diode laser (TEC-500-1060-030, Sacher Lasertechnik) with a custom-tuned wavelength range of 1000–1085 nm, linewidth of 10 kHz, and maximum tuning speed of 200 nm/s. This seed laser is linearly polarized, and outputs through a PM980 fiber. The seed laser is delivered to a polarization-maintained (PM) fiber-coupled $LiNbO_3$ electro-optic phase modulator (NIR-MPX-LN-02, Exail Technologies), driven by a radio-frequency amplifier seeded with a radio frequency sinusoidal source at 700 MHz oscillation frequency. The phase modulator is followed with a SOA intensity modulator (SOM-Shape-H1020, Aerodiode), to modulate the CW seed laser into a nanosecond pulse train operating at 1 kHz repetition rate.

The amplification module comprises three-stage Yb-doped fiber pre-amplifiers and a power amplifier, pumped by 976 nm laser diodes operating in pulsed mode (150–200 μs pulse width). The first two stages use PM Yb-doped single-mode fibers (PM-YSF-HI-HP, Coherent; mode-field diameter 7.5 μm at 1060 nm) with lengths of 6 m and 3 m, respectively, to optimize gain bandwidth and flatness. PM980 fiber-coupled acousto-optic modulators (AOMs, CSRayzer Optical Technology) were spliced after each amplification stage to suppress ASE and act as broadband isolators (1000–1085 nm) with damage thresholds >1 kW for nanosecond pulses. To enable the isolator function, additional PM980 fiber is spliced to the output side of each AOM, to ensure that the AOM switches to the 'off' state when the backward propagation pulses arrive. The bandwidth of the AOM is 200 MHz.

The 3$^{rd}$ stage pre-amplifier adopts a large-mode-field single-mode fiber (PM-YDF-HI-8_NF, Coherent), followed by a broadband fiber-coupled linear polarizer (1060 ± 50 nm) to maintain a polarization extinction ratio of 20 dB and suppress ASE. The power amplifier adopts a PM large-mode-area Yb-doped fiber (PLMA-YDF-25/250-UF, Coherent). The fiber output of the power amplifier is terminated with an end-cap, with a coating to limit the back-reflection to <0.1% at both signal (~1 μm) and pump (976 nm) wavelengths. To suppress the higher-order transverse modes, the fiber is coiled with a 10 cm diameter. The 3$^{rd}$ and power stage amplifiers employed a relatively short Yb-doped fiber length of 2 m to mitigate SBS. Mode-field-adaptors (Advanced Fiber Resources) are used to mitigate the coupling loss between different types of fibers. A trigger-delay electric unit is used to synchronize all pulsed 976 nm diodes, SOA, and two AOMs.

Four broadband optical circulators (1064±30 nm, Advanced Fiber Resources) are placed before each amplifier stage to isolate and monitor back-reflections. No isolator/circulator is used after the power amplifier due to high pulse energy and broad wavelength range. Circulators preceding the third and power amplification stages have a damage threshold of 1 kW peak power for few nanosecond pulses. The backward propagation Brillouin lasers are monitored through detecting the pulse profiles from the 3$^{rd}$ output port of each circulator with photodetectors (DET08C, Thorlabs) connected to an oscilloscope (MSO64B, Tektronix). Pump diode parameters were adjusted to keep the average Brillouin pulse peak below a certain value (see Fig. S7), which corresponds to 10% of the isolator damage threshold.

The wavelength tuning range of the pump is 1034 to 1085 nm, limited by circulator bandwidth (1064 ± 30 nm) and seed laser tuning capability (1000–1085 nm). Since SBS thresholds vary with wavelength due to ASE spectral dependence, re-optimization of pump diode parameters is required at each wavelength. Given this fact, a parameter database was established for practical operation, providing reference settings across the wavelength tuning range of 1034-1085 nm.

**Method 4: Wavelength stability measurement of Raman laser**

The wavelength stability of the 1228 nm Raman laser was evaluated by measuring the frequency difference between the Raman and pump lines using a near-infrared Fabry–Perot interferometer (SA210-8B, Thorlabs). This interferometer provides a FSR of 10 GHz and a resolution of 67 MHz over the 820–1275 nm range, allowing simultaneous measurement of both the pump (1050 nm) and Raman (1228 nm) signals. Interferometric traces were recorded every 5 min over a 1 h period (see Fig. S5g). Each trace contains four peaks, labelled as $P_1$, $P_2$, $P_3$, and $P_4$ in chronological order. $P_1$ and $P_3$ are from the pump line, and $P_2$ and $P_4$ from the Raman line (see Fig. S5g). The temporal center of each peak was calculated as $T_m = \frac{\sum_{t1}^{t2} t \cdot y(t)}{\sum_{t1}^{t2} y(t)}$, where $y(t)$ is the interferometric voltage signal, $t1$ and $t2$ are the left and right time boundaries of the selected interferometric envelop with peak $P_m$. The frequency difference shown in Fig. 3d was computed as $\frac{T_2 - T_1}{T_4 - T_2}$ for the black-color data and $\frac{T_4 - T_3}{T_4 - T_2}$ for the red-color data.

## Data Availability

All data generated in this work is publicly available through Figshare.

## Code Availability

No code is involved in this work.

## References

1. Jin, Y. *et al*. Long wavelength mid-infrared multi-gases spectroscopy using tunable single-mode slot waveguide quantum cascade laser. Opt. Express 31, 27543–27552 (2023).
2. Phillips, M. C. *et al*. Design and performance of a sensor system for detection of multiple chemicals using an external cavity quantum cascade laser. in Proc. SPIE vol. 7608 76080D (2010).
3. Phillips, M. C. *et al*. Real-time trace gas sensing of fluorocarbons using a swept-wavelength external cavity quantum cascade laser. Analyst 139, 2047–2056 (2014).
4. Zhang, C. *et al*. Multi-spectral optoacoustic microscopy driven by gas-filled hollow-core fiber laser

pulses. Biomed. Opt. Express 16, 3337–3348 (2025).

5. Shi, J. *et al*. High-resolution, high-contrast mid-infrared imaging of fresh biological samples with ultraviolet-localized photoacoustic microscopy. Nat. Photonics 13, 609–615 (2019).

6. Wang, X. *et al*. High-energy single-longitudinal-mode mid-infrared optical parametric amplifier seeded with sheet optical parametric oscillator. AIP Adv. 11, 065104 (2021).

7. Saikawa, J., Fujii, M., Ishizuki, H. & Taira, T. 52 mJ narrow-bandwidth degenerated optical parametric system with a large-aperture periodically poled MgO:LiNbO$_3$ device. Opt. Lett. 31, 3149–3151 (2006).

8. Ricciardi, I. *et al*. A narrow-linewidth optical parametric oscillator for mid-infrared high-resolution spectroscopy. Mol. Phys. 110, 2103–2109 (2012).

9. Saikawa, J., Miyazaki, M., Fujii, M., Ishizuki, H. & Taira, T. High-energy, broadly tunable, narrow-bandwidth mid-infrared optical parametric system pumped by quasi-phase-matched devices. Opt. Lett. 33, 1699–1701 (2008).

10. Vodopyanov, K. L., Makasyuk, I. & Schunemann, P. G. Grating tunable 4 - 14 μm GaAs optical parametric oscillator pumped at 3 μm. Opt. Express 22, 4131–4136 (2014).

11. He, Y. *et al*. High energy and tunable mid-infrared source based on BaGa4Se$_7$ crystal by single-pass difference-frequency generation. Opt. Express 27, 9241–9249 (2019).

12. Hu, S. *et al*. High-conversion-efficiency tunable mid-infrared BaGa4Se$_7$ optical parametric oscillator pumped by a 2.79 μm laser. Opt. Lett. 44, 2201–2203 (2019).

13. Hugi, A., Maulini, R. & Faist, J. External cavity quantum cascade laser. Semicond. Sci. Technol. 25, 083001 (2010).

14. Hugi, A. *et al*. External cavity quantum cascade laser tunable from 7.6 to 11.4 μm. Appl. Phys. Lett. 95, 061103 (2009).

15. www.toptica.com/products/tunable-diode-lasers/ecdl-dfb-lasers/ctl/

16. Sergachev, I. *et al*. Gain-guided broad area quantum cascade lasers emitting 23.5 W peak power at room temperature. Opt. Express 24, 19063–19071 (2016).

17. Pleitez, M. A. *et al*. Label-free metabolic imaging by mid-infrared optoacoustic microscopy in living cells. Nat. Biotechnol. 38, 293–296 (2020).

18. Park, N., Dawson, J. W., Vahala, K. J. & Miller, C. All fiber, low threshold, widely tunable single-frequency, erbium-doped fiber ring laser with a tandem fiber Fabry-Perot filter. Appl. Phys. Lett. 59, 2369–2371 (1991).

19. Wang, K. *et al*. Widely tunable ytterbium-doped single-frequency all-fiber laser. Opt. Laser Technol. 128, 106242 (2020).

20. Lu, B. *et al*. High-stability broadband wavelength-tunable single-frequency ytterbium-doped all-fiber compound ring cavity. IEEE Photonics J. 9, 1–8 (2017).

21. Honzatko, P., Baravets, Y. & Myakalwar, A. K. Single-frequency fiber laser based on a fiber ring resonator filter tunable in a broad range from 1023 nm to 1107 nm. Opt. Lett. 43, 1339–1342 (2018).

22. Wan, Y. *et al*. Broadband high-gain Yb: YAG crystal-derived silica fiber for low noise tunable single-frequency fiber laser. Opt. Express 30, 18692–18702 (2022).

23. Fan, Z. *et al*. High-power, widely wavelength-tunable, single-frequency pulsed fiber master oscillator power amplifier at 2.8 μm. High Power Laser Sci. Eng. 13, e26 (2025).

24. Zhang, Y. *et al*. Watt-level single-frequency lasing enabled by an $Er^{3+}$-doped fluoride fiber ring oscillator at 2.79 μm. Opt. Lett. 50, 6249–6252 (2025).

25. Poletti, F. Nested antiresonant nodeless hollow core fiber. Opt. Express 22, 23807–23828 (2014).

26. Gao, S. *et al*. Hollow-core conjoined-tube negative-curvature fibre with ultralow loss. Nat. Commun. 9, 2828 (2018).

27. Petrovich, M. *et al*. Broadband optical fibre with an attenuation lower than 0.1 decibel per kilometre. Nat. Photonics 19, 1203–1208 (2025).

28. Gao, S. F. *et al*. From Raman frequency combs to supercontinuum generation in nitrogen-filled hollow-core anti-resonant fiber. Laser Photon. Rev. 16, 2100426 (2022).

29. Cooper, M. A. *et al*. 2.2 kW single-mode narrow-linewidth laser delivery through a hollow-core fiber. Optica 10, 1253–1259 (2023).

30. Mulvad, H. C. H. *et al*. Kilowatt-average-power single-mode laser light transmission over kilometre-scale hollow-core fibre. Nat. Photonics 16, 448–453 (2022).

31. Shi, J. *et al*. All-fiber highly efficient delivery of 2 kW laser over 2.45 km hollow-core fiber. Nat. Commun. 16, 8965 (2025).

32. Wang, Y. *et al*. Synthesizing gas-filled anti-resonant hollow-core fiber Raman lines enables access to the molecular fingerprint region. Nat. Commun. 15, 9427 (2024).

33. Wang, Z., Gu, B., Chen, Y., Li, Z. & Xi, X. Demonstration of a 150-kW-peak-power, 2-GHz-linewidth, 1.9 μm fiber gas Raman source. Appl. Opt. 56, 7657–7661 (2017).

34. Chen, Y., Wang, Z., Gu, B., Yu, F. & Lu, Q. Achieving a 1.5 μm fiber gas Raman laser source with about 400 kW of peak power and a 6.3 GHz linewidth. Opt. Lett. 41, 5118–5121 (2016).

35. Li, Y., Chen, Z., Zhu, X., Wu, D. & Yu, F. 25 W continuous-wave fiber gas Raman laser at 1.9 μm wavelength based on low-loss anti-resonant hollow-core fibers. Chin. Opt. Lett., 23, 91403 (2025).

36. Li, Z., Huang, W., Cui, Y., Gu, B. & Wang, Z. High-efficiency, high peak-power, narrow linewidth 1.9 μm fiber gas Raman amplifier. J. Lightwave Technol. 36, 3700–3706 (2018).

37. Wang, Y. *et al*. High pulse energy and quantum efficiency mid-infrared gas Raman fiber laser targeting $CO_2$ absorption at 4.2 μm. Opt. Lett. 45, 1938–1941 (2020).

38. Li, Z., Huang, W., Cui, Y. & Wang, Z. Efficient mid-infrared cascade Raman source in methane-filled

hollow-core fibers operating at 2.8 μm. Opt. Lett. 43, 4671–4674 (2018).

39. Astapovich, M. S. et al. Watt-level nanosecond 4.42 μm Raman laser based on silica fiber. IEEE Photonics Technol. Lett. 31, 78–81 (2019).

40. Goel, C., Li, H., Abu Hassan, M. R., Chang, W. & Yoo, S. Anti-resonant hollow-core fiber fusion spliced to laser gain fiber for high-power beam delivery. Opt. Lett. 46, 4374–4377 (2021).

41. Wang, C., Yu, R., Xiong, C., Zhu, J. & Xiao, L. Ultralow-loss fusion splicing between antiresonant hollow-core fibers and antireflection-coated single-mode fibers with low return loss. Opt. Lett. 48, 1120–1123 (2023).

42. Huang, W. *et al*. Towards all-fiber structure pulsed mid-infrared laser by gas-filled hollow-core fibers. Chin. Opt. Lett. 17, 91402 (2019).

43. Song, W., Wen, Y., Zhang, Q., Zhang, X. & Wang, P. All-fiber-structure high-power mid-infrared gas-filled hollow-core-fiber amplified spontaneous emission source. Photonics Res. 13, 1137–1147 (2025).

44. Lv, G. *et al*. All-fiber structure tunable mid-infrared light source based on acetylene-filled hollow-core fibers. Opt. Express 33, 49353–49361 (2025).

45. Zhou, Z. *et al*. Towards high-power mid-IR light source tunable from 3.8 to 4.5 μm by HBr-filled hollow-core silica fibres. Light Sci. Appl. 11, 15 (2022).

46. Zhou, Z. *et al*. Nanosecond fiber laser step-tunable from 3.87 to 4.5 μm in HBr-filled hollow-core silica fibers. J. Lightwave Technol. 41, 333–340 (2023).

47. Shi, J. *et al*. Mid-infrared pulsed fiber laser source at 4.3 μm based on a $CO_2$-filled anti-resonant hollow-core silica fiber. Opt. Express 32, 43033–43045 (2024).

48. Zhou, Z. *et al*. High-power tunable mid-infrared fiber gas laser source by acetylene-filled hollow-core fibers. Opt. Express 26, 19144–19153 (2018).

49. Huang, W. *et al*. Tunable fiber gas Raman laser of 6 w at 2.9 μm by deuterium-filled hollow-core fiber. IEEE J. Sel. Top. Quantum Electron. 30, 1–7 (2024).

50. Li, H., Huang, W., Cui, Y., Zhou, Z. & Wang, Z. 3 W tunable 1.65 μm fiber gas Raman laser in $D_2$-filled hollow-core photonic crystal fibers. Opt. Laser Technol. 132, 106474 (2020).

51. Hong, L. *et al*. High energy and narrow linewidth $N_2$-filled hollow-core fiber laser at 1.4 μm. J. Light. Technol 1–5 (2024).

52. Wang, Y., Schiess, O. T. S., Amezcua-Correa, R. & Markos, C. $CO_2$-based hollow-core fiber Raman laser with high-pulse energy at 1.95 μm. Opt. Lett. 46, 5133–5136 (2021).

53. Hanna, D., Pointer, D. & Pratt, D. Stimulated Raman scattering of picosecond light pulses in hydrogen, deuterium, and methane. IEEE J. Quantum. Electron. 22, 332–336 (1986).

54. Liu, Y. *et al*. >1 kW all-fiberized narrow-linewidth polarization-maintained fiber amplifiers with wavelength spanning from 1065 to 1090 nm. Appl. Opt. 56, 4213–4218 (2017).20

55. Ma, P. *et al*. 1.89 kW all-fiberized and polarization-maintained amplifiers with narrow linewidth and near-diffraction-limited beam quality. Opt. Express 24, 4187–4195 (2016).

56. Ionov, P. I. & Rose, T. S. SBS reduction in nanosecond fiber amplifiers by frequency chirping. Opt. Express 24, 13763–13777 (2016).

57. Pouillaude, J. *et al*. Development of an all-fiber spliced laser system using stimulated Brillouin scattering mitigation techniques and achieving 50 ns, 10 kW peak power for lidar applications. Opt. Lett. 50, 1743–1746 (2025).

58. Ran, Y. *et al*. 293 W, GHz narrow-linewidth, polarization maintaining nanosecond fiber amplifier with SBS suppression employing simultaneous phase and intensity modulation. Opt. Express 23, 25896–25905 (2015).

59. Zeringue, C., Dajani, I., Naderi, S., Moore, G. T. & Robin, C. A theoretical study of transient stimulated Brillouin scattering in optical fibers seeded with phase-modulated light. Opt. Express 20, 21196–21213 (2012).

60. Huang, X., Liang, S., Xu, L., Richardson, D. J. & Jung, Y. Wideband (13.7 THz) gain-flattened Yb-doped fiber amplifier for telecommunication applications. IEEE Photonics Technol. Lett. 36, 791–794 (2024).

61. https://www.nktphotonics.com/products/optical-fibers-and-modules/

62. Anderson, B., Flores, A., Holten, R. & Dajani, I. Comparison of phase modulation schemes for coherently combined fiber amplifiers. Opt. Express 23, 27046–27060 (2015).

63. Panbiharwala, Y. *et al*. Stimulated Brillouin scattering mitigation using optimized phase modulation waveforms in high power narrow linewidth Yb-doped fiber amplifiers. Opt. Express 29, 17183–17200 (2021).

64. Herring, G. C., Dyer, M. J. & Bischel, W. K. Temperature and density dependence of the linewidths and line shifts of the rotational Raman lines in $N_2$ and $H_2$. Phys Rev A (Coll Park) 34, 1944–1951 (1986).

65. Phillips, M. C. *et al*. Design and performance of a sensor system for detection of multiple chemicals using an external cavity quantum cascade laser. in Proc. SPIE vol. 7608 76080D (2010).

## Acknowledgements

This work is supported by LUNDBECK Fonden (Grant No. R346-2020-1924, R276-2018-869), and VILLUM Fonden (Grant No. 36063, Grant No. 40964).

## Ethics declarations

Competing interests

**Supplementary Information for "Tunable, high pulse energy and narrow linewidth gas-filled fiber laser across near- and mid-infrared"**

Yazhou Wang, Marcello Meneghetti, Manoj K. Dasa, Timothy Bate, J. E. Antonio-Lopez, Ian A. Davidson, Qiang Fu, Jaroslaw Rzegocki, Gregory T. Jasion, Francesco Poletti, Rodrigo Amezcua-Correa, and Christos Markos

# SUPPLEMENTARY NOTE 1. Characteristics of ARHCFs

Five anti-resonant hollow-core fibers (ARHCFs) are employed in this work, denoted as ARHCF-1 to ARHCF-5. Their lengths used for gas-filled Raman frequency conversion are 5 m, 5 m, 10 m, 5 m, and 10 m, respectively. Figure S8a–S8e summarizes their structural and transmission characteristics. ARHCF 1, 2, and 4 were designed, fabricated, and characterized at CREOL, The College of Optics and Photonics, University of Central Florida. ARHCF 3 and 5 were designed, fabricated, and characterized at the Optoelectronics Research Centre, University of Southampton.

ARHCF-1 (Fig. S8a): This fiber has a core diameter of 31 µm, surrounded by a cladding of seven capillaries with a wall thickness of 290 nm and a diameter of 16.8 µm. It exhibits a low-loss transmission window spanning ~ 800 to 1600 nm, which is a prerequisite for the efficient Raman frequency conversion in the near-infrared region (limited to wavelengths <1.6 µm).

ARHCF-2 (Fig. S8b): ARHCF-2 is the same as the $2^{nd}$ stage ARHCF used in Ref. [1]. Its characteristics are reproduced here for completeness. The cladding comprises seven nested silica capillaries, forming a circular hollow-core region with 82 µm diameter. The external capillaries in the cladding have a diameter and wall thickness of 40.3 µm and 987 nm, and internal capillaries have a diameter and wall thickness of 13.6 µm and 1.37 µm, respectively. This structure enables two transmission windows: ~1.2–1.6 µm and ~3–4.6 µm. A loss bump near 1.4 µm, caused by the internal capillaries, compromises Raman pulse energy to nanojoules around 3.3 µm wavelength (see Fig. 1e).

Experimental loss measurement of ARHCF-2 via the cut-back method was not performed, as the available ARHCF-2 sample is of limited length and reserved for other experiments.

ARHCF-3 (Fig. S8c): This fiber features a 39 µm core surrounded by five nested capillaries with a wall thickness of 620 nm. It offers a broad transmission window from ~1.45 to 3 µm, supporting $H_2$-based cascaded rotational Raman frequency conversion (Fig. 1b). Experimental loss measurements in Fig. S8c are limited to 1.6–2.4 µm due to weak light signal beyond this range during the cut-back measurements from 2 km to 20 m.

ARHCF-4 (Fig. S8d): ARHCF-4, identical to the fiber in Ref. [2], has a 23 µm core diameter and a cladding of five nested capillaries with a wall thickness of 780 nm. It provides a low-loss transmission window between ~0.9 and 1.3 µm, followed by a high-loss resonance peak beyond 1.3 µm. This loss profile boosts the pulse energy of Raman lasers at 1207–1277 nm based on the $1^{st}$ order $CO_2$ vibrational Stokes, since it blocks high-order vibrational Raman Stokes.

ARHCF-5 (Fig. S8e): This fiber comprises five nested capillaries forming a 32 µm core, with a wall thickness of 394 nm. It exhibits a low-loss transmission window from 1.1 to 2.7 µm, enabling Q(1) vibrational Raman frequency conversion of $H_2$ from the ~1.2 µm pump region to mid-infrared region (see Fig. 3a). The measured loss in Fig. S8e is limited to the wavelength range of 1.2 to 2.4 µm; beyond this range, the light signal was too weak for detection during cut-back measurements from 511 m to 10 m.

## SUPPLEMENTARY NOTE 2. Attenuation of Raman lines in ambient air

Narrow-linewidth Raman laser can experience significant attenuation during propagation in ambient air when its wavelength coincides with one of the strong mid-infrared absorption lines of atmospheric gases. A notable example occurs near 4.23 µm, which is located at the center of the fundamental absorption band of $CO_2$. To quantify this effect, we measured the power of a collimated Raman beam at 4.23 µm as a function of propagation distance, following the method described in Ref. [1]. By properly tuning the Raman line at around 4.23 µm, an attenuation coefficient of 8.1 $m^{-1}$ was observed, as shown in Fig. S9a. For comparison, Fig. S9b shows the simulated attenuation coefficient of $CO_2$ at an ambient concentration of 400 ppm, where the peak attenuation is ~12 m⁻¹. This indicates that the experimentally measured attenuation coefficient approaches the theoretical maximum for $CO_2$ absorption.

Based on these results, the Raman laser wavelengths used for linewidth measurements in this work were carefully tuned to avoid absorption features of $CO_2$, water vapor, and other atmospheric constituents.

## SUPPLEMENTARY NOTE 3. Theoretical wavelength tuning speed of the proposed Raman laser

The wavelength tuning speed of a Raman Stokes line stems from the relationship between the Raman Stokes wavelength ($\lambda_R$) and the pump wavelength ($\lambda_P$), which is given by:

$$\lambda_R = \frac{1}{\frac{1}{\lambda_P} - \Omega} \quad (1)$$

where $\Omega$ is the Raman shift coefficient. Differentiating Eq. (1) with respect to time yields:

$$\frac{d\lambda_R}{dt} = \frac{1}{(1-\Omega\lambda_P)^2} \frac{d\lambda_P}{dt} \quad (2)$$

where $\frac{d\lambda_R}{dt}$ and $\frac{d\lambda_P}{dt}$ represent the tuning speeds of the Raman Stokes and pump wavelengths, respectively.

Figure S10 illustrates the calculated Raman wavelength tuning speed as a function of Raman wavelength. The tuning speed accelerates toward longer wavelengths. In this calculation, the pump tuning speed is set to 200 nm/s, equal to the maximum tuning speed of the external-cavity diode laser seed specified by the supplier. The pump wavelength is fixed at 1060 nm, and $\Omega$ is varied from 0-7450 cm⁻¹, to scan the Raman wavelength.

Figure S10 also shows the theoretical resolution of the Raman wavelength tuning, assuming a pump wavelength tuning resolution of 35 fm based on the seed laser specifications. The calculated curve indicates that acceleration of the tuning speed comes at the expense of compromising tuning resolution.

**Fig. S1: Complete spectra corresponding to Fig. 1b in the wavelength range 0.5–5 μm.**

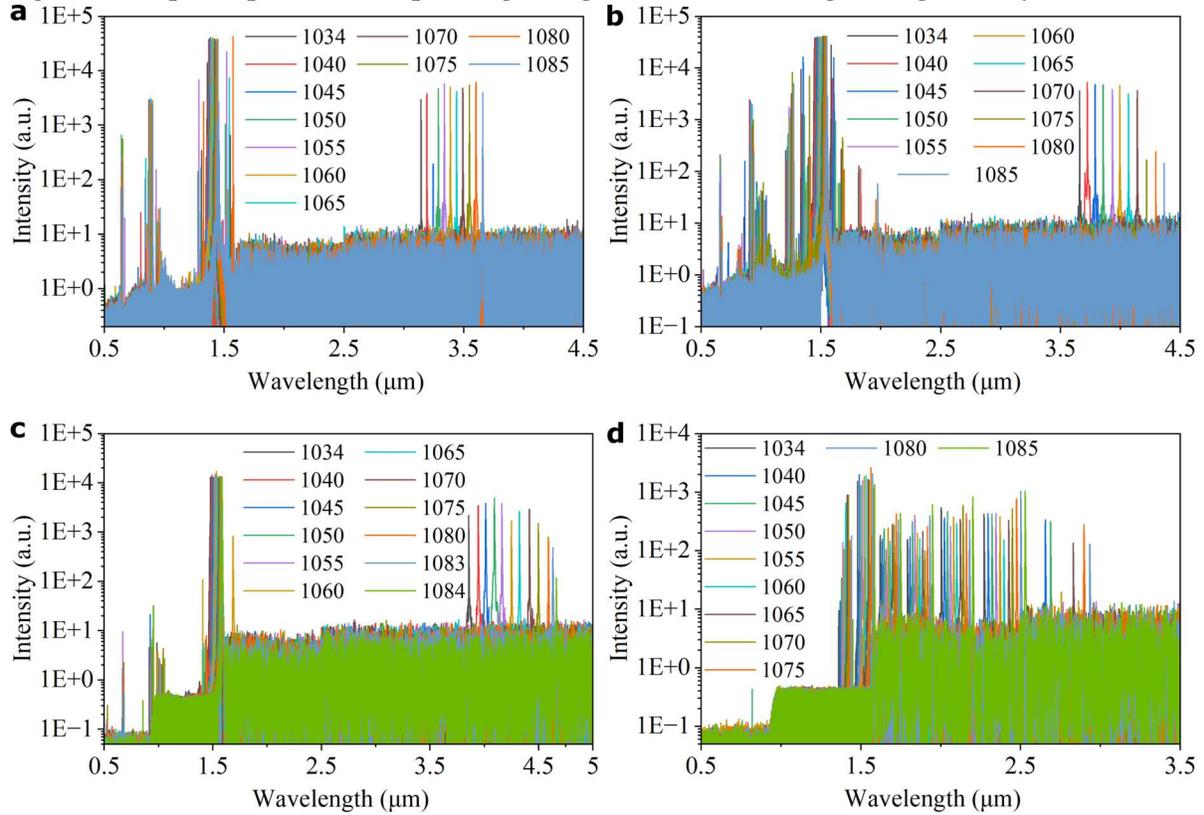

**a-c** Spectra generated in the cascaded ARHCF-1 and ARHCF-2 configuration. ARHCF-2 is filled with $H_2$ at 30 bar, while ARHCF-1 is filled with $N_2$ at 15 bar in **a**, $CO_2$ at 2 bar in **b**, and $CH_4$ at 2 bar in **c**. **d** Spectra generated from cascaded $CH_4$-filled ARHCF-1 and $H_2$-filled ARHCF-3. The legends indicate the pump wavelength (in nm) used for each measurement. Pump spectral lines in 1034-1085 nm are not shown, because they are blocked at the output end of the 1st stage ARHCF using a 1.3 μm long pass filter. Spectra were recorded using a spectrum meter (Spectro 320, Instrument Systems) with a resolution of 0.1 nm (<632 nm), 0.2 nm (632–2000 nm), and 0.4 nm (>2000 nm). A multimode $InF_3$ fiber patch cable (MF11L1AR1, Thorlabs) was used for coupling lasers into the spectrum meter.

**Fig. S2: Characterization of gas-filled ARHCF-1 Raman lasers.**

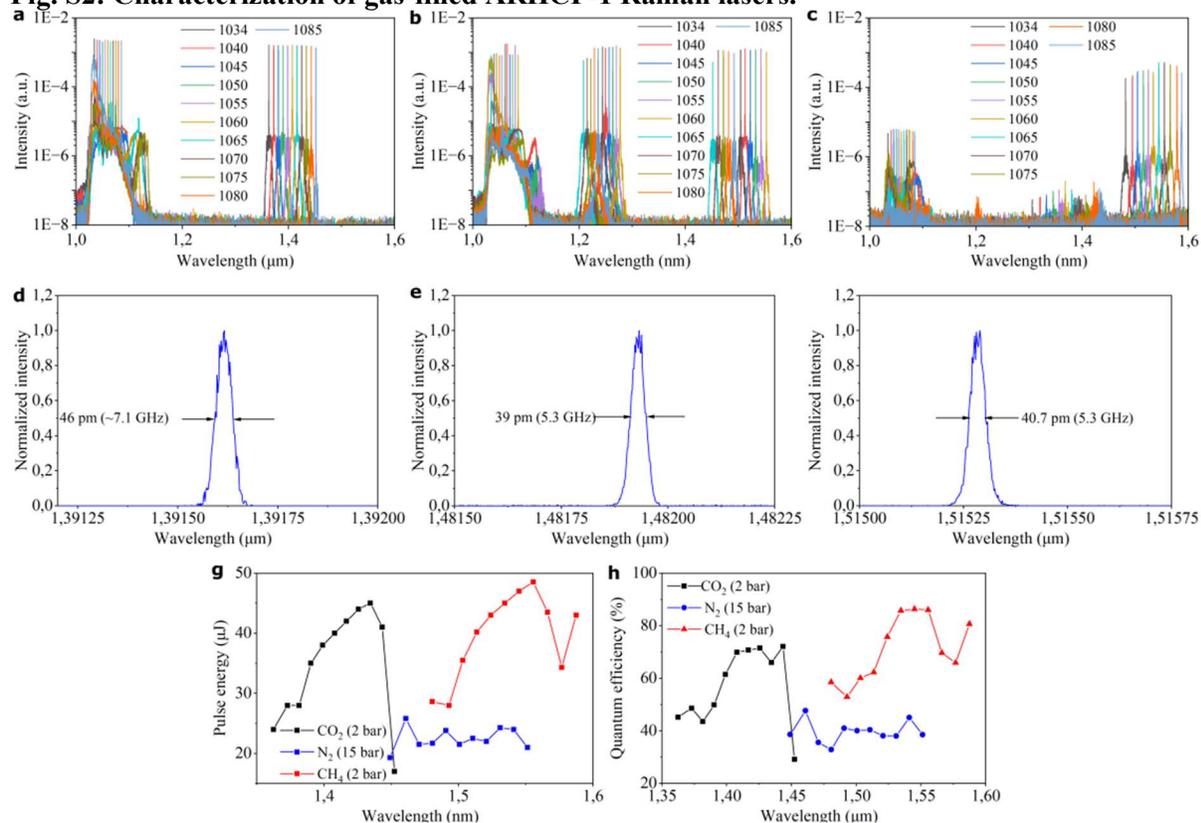

**a-c** Measured spectra when ARHCF-1 is filled with $N_2$ at 15 bar in **a**, $CO_2$ at 2 bar in **b**, and $CH_4$ at 2 bar in **c**. **d-f** Linewidth of representative Raman line from panels **a**, **b**, and **c**, respectively. **g**, **h** Measured pulse energy and quantum efficiency of ARHCF-1 Raman lasers, based on $N_2$, $CO_2$, and $CH_4$. All spectra were recorded using a spectrum analyzer (ANDO AQ6317B, AssetRelay) with 10 pm resolution. Laser output was coupled into the spectrum analyzer via an endless single-mode fiber (ESM-12B, Thorlabs) patch cable, with FC/PC termination. Pulse energy was measured using an energy meter (PE9-ES-C, Ophir Optronics).

**Fig. S3: Linewidth measurement of mid-infrared Raman lines.**

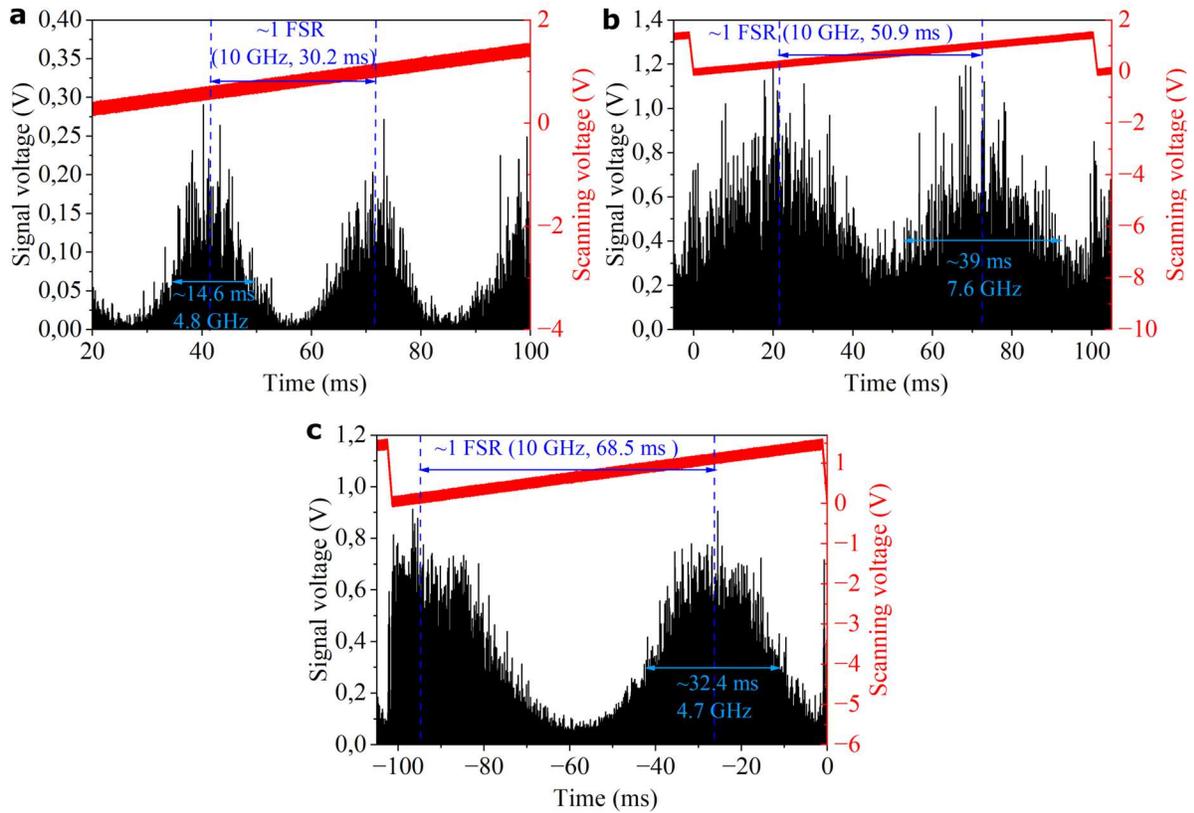

**a** Fabry–Perot interferometric trace of a 2.68 μm Raman line generated in cascaded $CH_4$-filled ARHCF-1 and $H_2$-filled ARHCF-3. **b** Trace of a 3.39 μm Raman line from cascaded $N_2$-filled ARHCF-1 and $H_2$-filled ARHCF-2. **c** Trace of a 3.99 μm Raman line from cascaded $CO_2$-filled ARHCF-1 and $H_2$-filled ARHCF-2. Panel **a** is measured using SA210-18C (Thorlabs), while panels **b** and **c** use SA210-30C-SP (Thorlabs). All Fabry–Perot interferometers have a free spectral range of 10 GHz and a resolution of 67 MHz in the specified wavelength range (see Method 2).

**Fig. S4: Average power and pulse energy of the pump laser with phase modulation.**

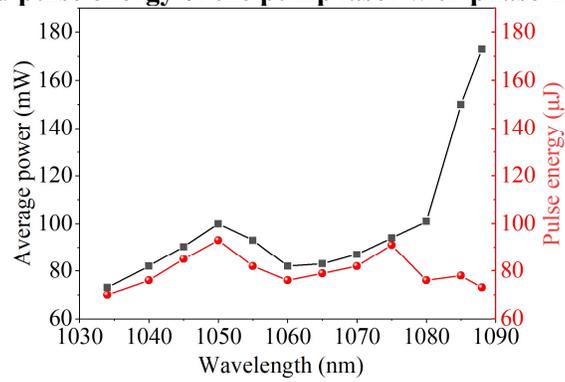

Average power was measured using a thermal power meter (S405C, Thorlabs). Pulse energy was measured using an energy meter (PE9-ES-C, Ophir Optronics) to exclude the influence of ASE.

## Fig. S5: Supplementary results related to Fig. 3.

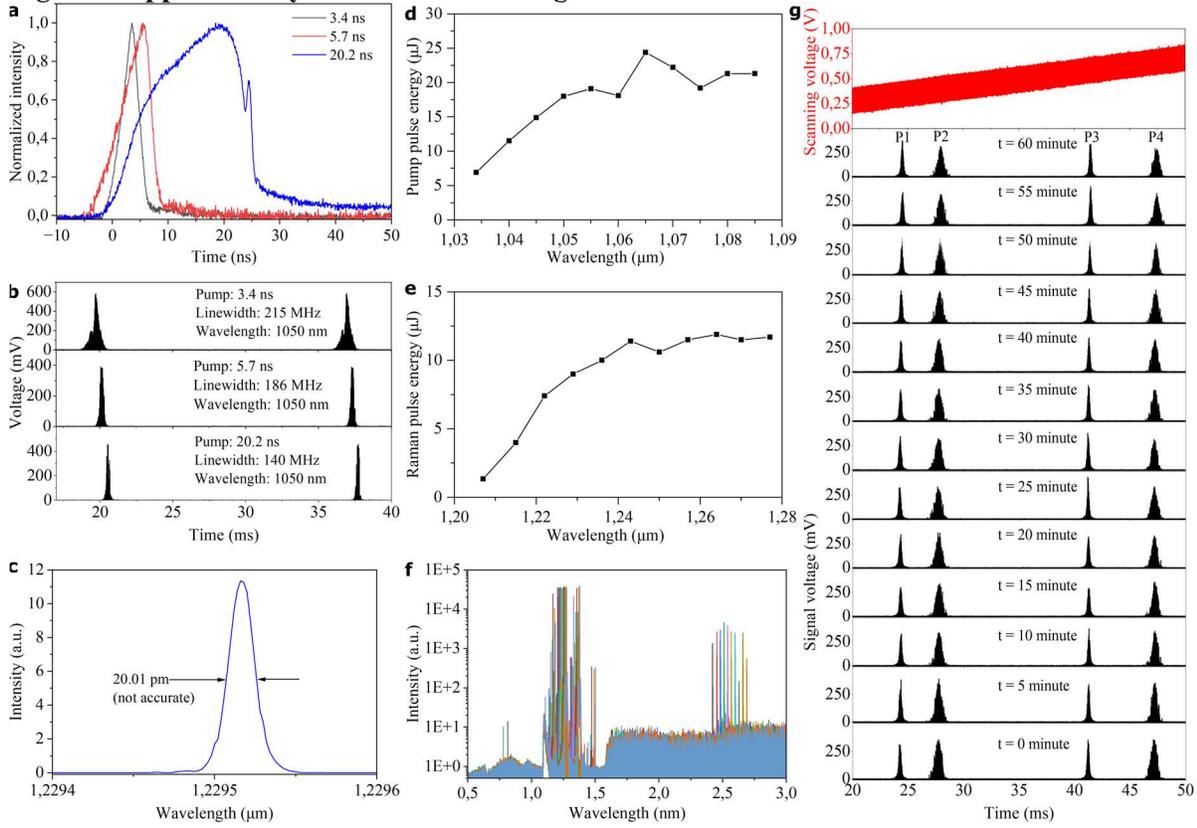

**a** Pump pulse profiles with durations of 3.4 ns, 5.7 ns, and 20.2 ns, measured using a 5 GHz bandwidth photodetector (DET08C, Thorlabs). **b** Linewidths of the pump laser at the three pulse durations shown in panel **a**, measured using a Fabry–Perot interferometer (SA210-8B, Thorlabs). **c** Optical spectrum of the $CO_2$-filled ARHCF-4 vibrational Raman Stokes line at 1.23 μm wavelength, recorded with an optical spectrum analyzer (ANDO AQ6317B, AssetRelay; resolution 10 pm). The measured linewidth of 20 pm is not accurate, due to the limited resolution of the spectrum analyzer. **d** Pulse energy of the wavelength tunable pump at 5.7 ns pulse duration. **e** Pulse energies of the 1st order vibrational Raman laser from the 1st stage $CO_2$-filled ARHCF-4, pumped with the energy shown in panel **d**. **f** Spectra of the wavelength-tunable Raman output from the 2nd stage $H_2$-filled ARHCF-5, measured using Spectro 320 (Instrument Systems). **g** Raw data used to calculate frequency stability in Fig. 3d.

**Fig. S6: Design of gas cells for sealing ARHCFs.**

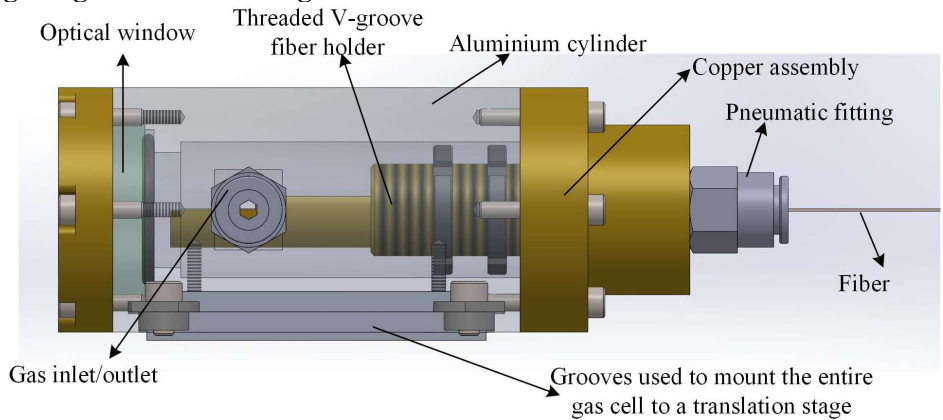

The gas cell consists of a hollow-core aluminum cylinder with an internal diameter of approximately 2 cm and length of 7 cm. One end of the cylindrical body is sealed with an optical window, while the opposite end is closed by a copper assembly that also features a hollow-core geometry. This copper assembly houses a threaded V-groove fiber holder, allowing precise adjustment of the holder's insertion depth within the cell. A lens can be mounted onto the fiber holder (inside the gas cell) via a standard lens tube, enabling controlled beam delivery. The distal end of the copper assembly interfaces with a gas hose containing the ARHCF. The ARHCF traverses the copper assembly and is mounted onto the V-groove fiber holder.

**Fig. S7: Backward-propagating SBS pulses generated in the power amplification stage of the pump laser.**

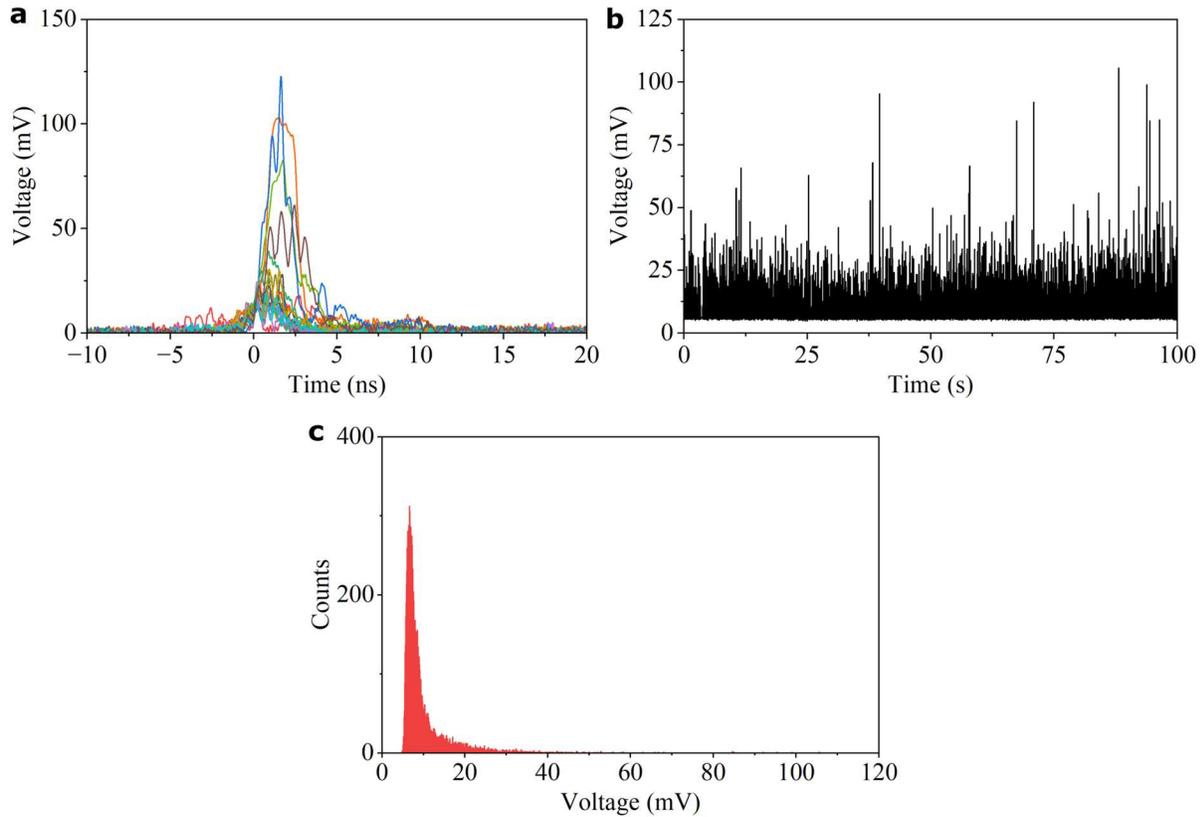

**a** Twenty typical recorded SBS pulse profiles at the conditions of 1060 nm wavelength and 76 µJ pulse energy (see Fig. 2d). **b** Peak voltage monitoring of SBS pulses over ~100 s. **c** Histogram of peak voltage values shown in panel **b**. SBS pulses were monitored using a photodetector (DET08C, Thorlabs) placed at the 3$^{rd}$ output port of the optical circulator preceding the power amplifier (see Fig. 2a). The average peak voltage, calculated with an oscilloscope, is always controlled to less than a value corresponding to ~100 W optical peak power — 10% of the damage threshold of the optical circulator.

**Fig. S8: Characterization of ARHCFs.**

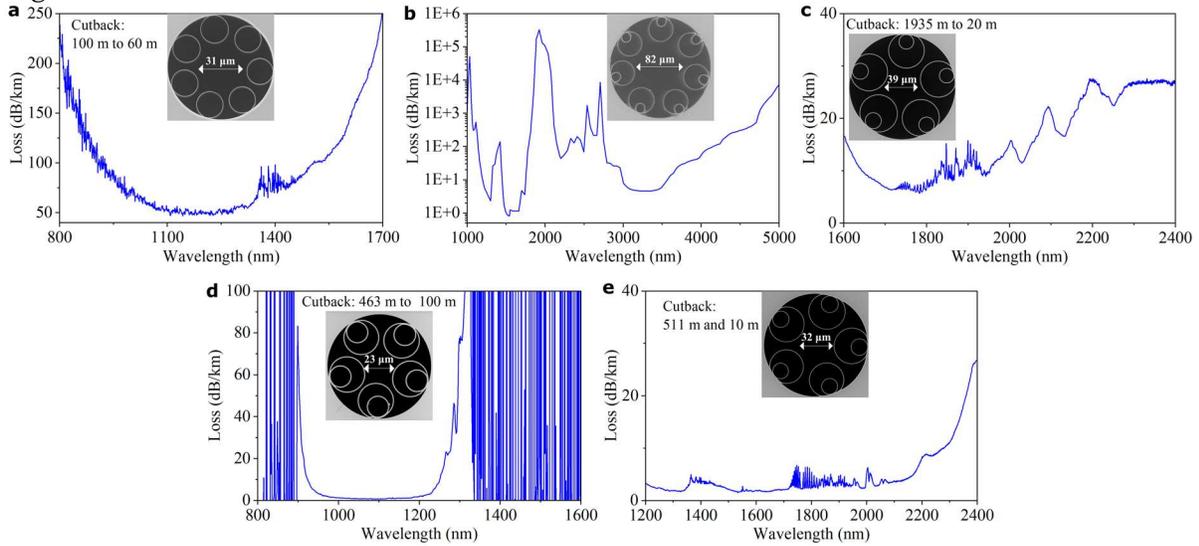

**a–e,** Characteristics of (**a**) ARHCF-1, (**b**) ARHCF-2, (**c**) ARHCF-3, (**d**) ARHCF-4, and (**e**) ARHCF-5. Insets show scanning electron microscope images of each fiber. Losses of ARHCF-1, ARHCF-3, ARHCF-4, and ARHCF-5 are experimentally measured using the cut-back method. The loss of ARHCF-2, identical to the 2$^{nd}$ stage ARHCF reported in Ref. [1], is simulated using the finite-element method (COMSOL). Experimental loss measurement for ARHCF-2 was not performed because the available sample length (5 m) is reserved for other experiments. ARHCF-4 corresponds to the fiber used in Ref. [2].

**Fig. S9: Attenuation of a 4.23 μm Raman line during propagation in ambient air.**

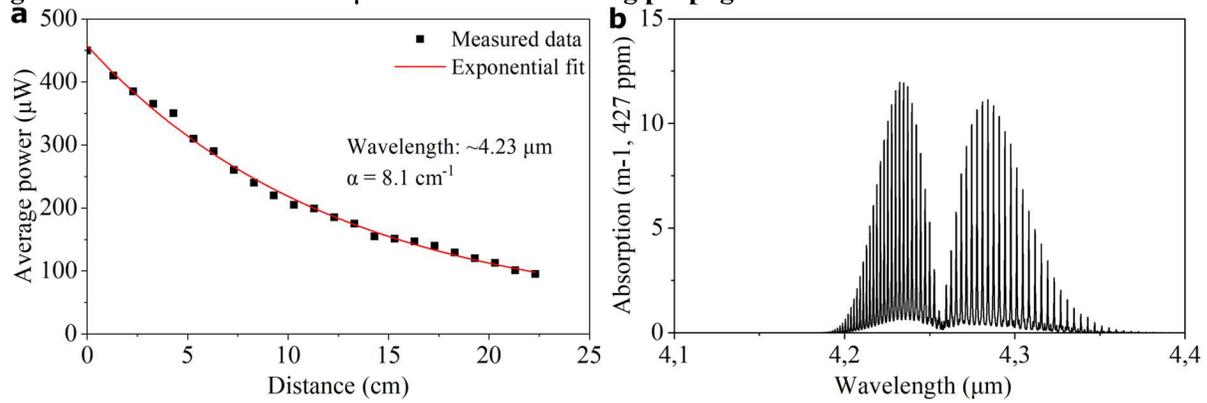

**a** Measured attenuation of the 4.23 μm Raman laser in ambient air. **b** Simulated $CO_2$ absorption spectrum assuming 400 ppm concentration in air, based on the HITRAN database (https://www.hitran.com/).

**Fig. S10: Theoretically calculated speed and resolution of Raman laser wavelength tuning.**

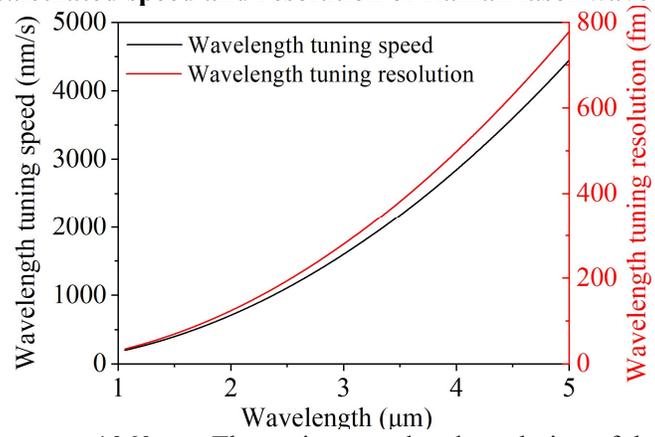

The pump wavelength was set to 1060 nm. The tuning speed and resolution of the pump are 200 nm/s and 35 fm, respectively, determined by the seed laser (NL-1060-0924-00402, Lasertechnik).

# Reference

1. Wang, Y. *et al*. Synthesizing gas-filled anti-resonant hollow-core fiber Raman lines enables access to the molecular fingerprint region. Nat. Commun. 15, 9427 (2024).

2. Cooper, M. A. *et al*. 2.2 kW single-mode narrow-linewidth laser delivery through a hollow-core fiber. Optica 10, 1253–1259 (2023).